# Strain and Twist Engineering of Interfacial Thermal Transport in Homo- and Hetero-Interfaces of Graphene and Hexagonal Boron Nitride

Wenwu Jiang[1,2], Huasong Qin[3*], Yilun Liu[3], Wengen Ouyang[1,2*],

Oded Hod[4*], Michael Urbakh[4]

[1]*Department of Engineering Mechanics, School of Civil Engineering, Wuhan University, Wuhan, Hubei 430072, China*

[2]*State Key Laboratory of Water Resources Engineering and Management, Wuhan University, Wuhan, Hubei 430072, China*

[3]*Laboratory for Multiscale Mechanics and Medical Science, SVLAB, School of Aerospace, Xi'an Jiaotong University, Xi'an 710049, China*

[4]*School of Chemistry and The Sackler Center for Computational Molecular and Materials Science, Tel Aviv University, Tel Aviv 6997801, Israel*

[*]Corresponding authors. E-mail: huasongqin@xjtu.edu.cn, w.g.ouyang@whu.edu.cn, odedhod@tauex.tau.ac.il

ABSTRACT

A dramatic difference between the vertical thermal conductance response of homogeneous and heterogeneous graphene/*h*-BN interfaces to external mechanical perturbations, is predicted. Homogeneous graphene and *h*-BN interfaces exhibit strong conductance reduction for both in-plane strain and interfacial twist. Conversely, the vertical thermal conductance of the heterogeneous graphene/*h*-BN junction is insensitive to twist deformations but shows significant increase or decrease under compressive or tensile strains, respectively. Our atomistic simulations predictions are rationalized by Fermi's golden rule and density of phonon modes analyses, indicating that vertical phonons and local stacking configurations have a central role in the interlayer heat transport behavior. A simple phenomenological model, based on local interlayer distance and stacking, captures well the dependence of vertical heat conductance on strain and twist deformations.

In the last decade, van der Waals (vdW) heterostructures have attracted vast attention due to their remarkable mechanical [1,2], magnetic [3,4], electronic [5-9], frictional [10-13], and thermal properties [14-20]. Of particular interest in electronics and spintronics is the heterogeneous graphene/hexagonal boron nitride (*h*-BN) interface that harnesses the unique electronic properties of graphene [21-25] and the insulating and isolating properties of *h*-BN. The relatively weak dispersive interlayer interactions between graphene and *h*-BN facilitate the fabrication of heterogeneous low-dimensional structures [26-30] that help alleviate the often undesirable semi-metallic nature of graphene [5,17,31-33].

As electronic and spintronic devices scale down and power density rises, managing and controlling heat dissipation becomes a critical challenge for guaranteeing device performance and reliability. In this respect, the integration of graphene and *h*-BN in such devices requires the regulation of their lateral and vertical thermal transport [34-43]. Twisted thermotics – a branch of twistronics, which examines the twist angle dependence of intra- and inter-layer thermal transport, may offer such control mechanisms [44-46]. This has been theoretically predicted and experimentally observed in various systems, including twisted multilayer graphene, *h*-BN, and transition metal dichalcogenides (TMDs) [42,47-58]. For homogeneous interfaces of graphene and *h*-BN, it was demonstrated that interlayer twist can serve as an effective knob for switching vertical heat transport [42,47,48,51,59]. This idea was later extended to the case of highly strained heterogeneous interfaces [43]. Here, we show that vertical thermal conductivity in unstrained graphene/*h*-BN heterojunctions, often encountered in experiments, is insensitive to twist deformations. Alternatively, we suggest in-plane strain as a stand-alone thermal transport control parameter in graphene/*h*-BN interfaces.

To that end, we performed fully-atomistic simulations of the thermal transport across homogeneous graphene and *h*-BN stacks, as well as graphene/*h*-BN heterostructures. The simulation model consists of 16-layered vdW stacks of various compositions [see FIG. 1(a)]. The effects of lateral strain and twist angle on the thermal transport were considered independently. Biaxial lateral strain was applied to one of the contacting materials within the stack [orange-colored layers in FIG. 1(a)]. Twisting was introduced by laterally rotating the entire sub-stack of one of the contacting materials [orange-colored layers in FIG. 1(a)] with respect to its counterpart (grey-colored layers). Once the constraint was applied, the entire stack was allowed to relax.

All simulations were conducted using three-dimensional periodic boundary conditions, while keeping the residual strain induced by the finite supercell dimensions below 0.012%. VdW interactions were described using dedicated anisotropic interlayer potentials (ILP) [10,60-70]. Intralayer interactions were described using the reactive empirical bond-order (REBO) potential [71] and the Tersoff potential [72] for graphene and *h*-BN, respectively. To simulate thermal transport, a temperature gradient was introduced by applying Langevin thermostats of different temperatures to the two layers [marked by the red and blue rectangles in FIG. 1(a)] that lie most remotely from the strained or twisted interface. This approach has been successfully used to calculate the vertical thermal transport of various vdW layered materials [42,49,53,73-76]. Specifically, good agreement was obtained between the calculated vertical thermal conductivities and experimental results for ~100-layers thick graphitic stacks [42]. Further details regarding modeling the periodic supercells, and the simulation protocol that was implemented within the LAMMPS package [77], are provided in Sec. 1 of the Supplemental Material (SM)[78-82].

We start by investigating the effects of the in-plane strain on the vertical thermal conductivity ($\kappa_\perp$) of homogeneous graphene and $h$-BN stacks. FIG. 1(b) presents $\kappa_\perp$ for 16-layered graphene and $h$-BN systems as a function of strain, $\varepsilon$, where positive (negative) values correspond to tensile (compressive) strain (see SM Secs. 1 and 2 for computational details [78]). For the unstrained homogeneous graphene and $h$-BN stacks we obtain $\kappa_\perp = 0.290 \pm 0.016$ and $0.559 \pm 0.026$ Wm$^{-1}$K$^{-1}$, respectively. Upon applying tensile or compressive strains a significant reduction in thermal transport is found. For graphene stacks, a strain of 3.1% (-3.8%) results in a decrease of ~87% (~45%) in $\kappa_\perp$ and for $h$-BN stacks a similar decrease of ~91% (~41%) is predicted. Notably, this strain-induced thermal conductivity switching capability compares well with $\kappa_\perp$ reduction obtained via twist deformations of homogeneous graphene and $h$-BN stacks (see Sec. 3 of SM [78]) [42,47,51].

A similar qualitative picture arises when calculating the interfacial thermal conductance (ITC, see Sec. 2 of SM for details [78]), which describes the heat flow at the strained (or twisted) interface within the stack. FIG. 1(c) and FIG. 1(d) present the ITC (normalized to its value at zero strain and twist angle $\langle \text{ITC}(\varepsilon, \theta)\rangle/\langle \text{ITC}(\varepsilon = 0, \theta = 0)\rangle$) for the strained homogeneous graphene and $h$-BN interfaces, respectively, as calculated using non-equilibrium molecular dynamics (NEMD) simulations. We find that applying pure tensile or compressive strain of $< \pm 1\%$ results in an ITC reduction of ~80%, thus demonstrating excellent switching capability.

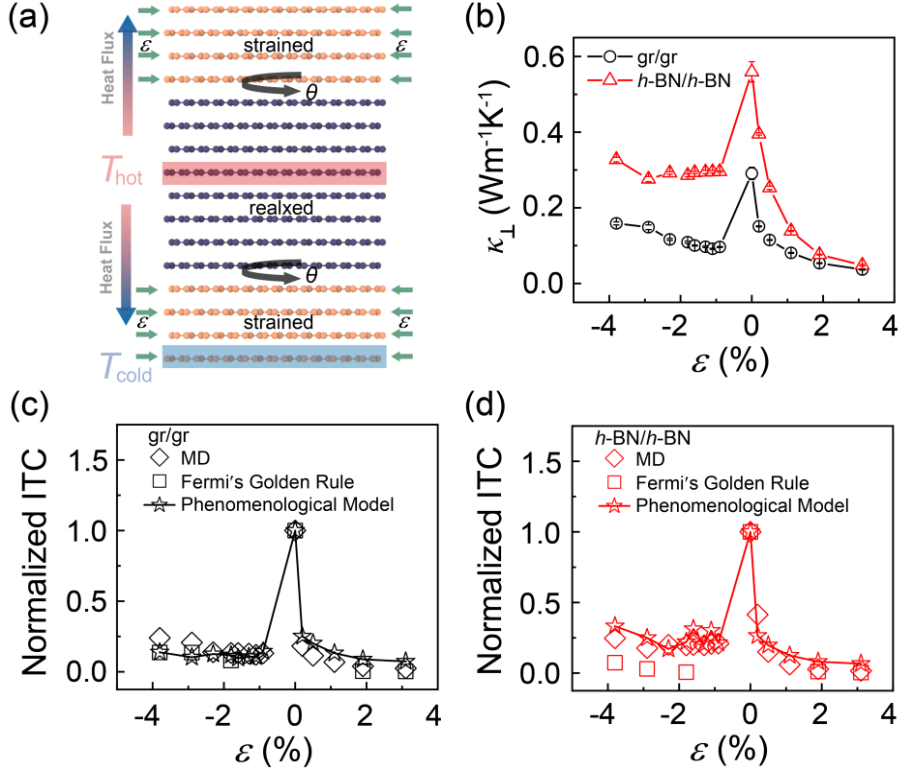

FIG. 1. (a) Schematic of the model system and setup for thermal transport simulations in homogeneous or heterogeneous van der Waals stacks. Two slabs (marked by gray and orange spheres) are strained or twisted relative to each other by an angle $\theta$. For homogeneous systems, both the gray and orange slabs correspond to the same material, either graphene or $h$-BN. For heterogeneous graphene/$h$-BN systems, the gray and orange slabs represent graphene and $h$-BN, respectively. The strain is exclusively applied to the orange slabs. Thermal gradients are introduced via Langevin thermostats, denoted by red (heat source) and blue (heat sink) rectangles. Periodic boundary conditions are applied in both the lateral and vertical directions. The latter results in a supercell including two twisted interfaces, across which heat flux flows in opposite directions (see vertical arrows). (b) Strain dependence of the vertical thermal conductivity of 16-layer homogeneous graphene (black circle) and $h$-BN (red triangle) stacks, respectively. (c) and (d) compare the calculated interfacial thermal conductance (ITC) obtained via NEMD simulations (black and red diamonds) with Fermi's golden rule (black and red squares) and phenomenological model (black and red stars) predictions for strained graphene and $h$-BN stacks, respectively. Here, all values are normalized with respect to the value obtained for the corresponding unstrained and aligned interface ($\varepsilon = 0, \theta = 0$).

An intuitive understanding of this behavior can be obtained by considering the degree of localization of the out-of-plane phonons, responsible for vertical heat transport. This can be achieved by considering bilayer graphene or bilayer $h$-BN, and block diagonalizing the Hessian matrix separately for the top and bottom layers (see details in Sec. 4 of SM [78] and in Ref. [42]). The resulting off-diagonal block matrix elements provide the couplings between phonon modes of the contacting materials, and their magnitude indicates how localized or delocalized these phonons modes are, residing either on each layer individually or spanning the entire bilayer system. Delocalized phonons are expected to carry heat across the interface, whereas localized phonons do not contribute to the overall heat transport. The normalized heat transfer rates obtained from the coupling matrix elements via Fermi's golden rule [see black and red rectangles of FIG. 1(c) and FIG. 1(d)] show good agreement with the ITC obtained from NEMD simulations (black and red diamonds) for both systems. This indicates that the dependence of the degree of phonon (de)localization on the induced deformation is the dominating factor determining the heat transport behaviors presented above.

A more quantitative description can be obtained by calculating the frequency resolved thermal conductance [see FIG. 2], $G(\omega)$, as a function of deformation strength using the spectral heat current (SHC) [83-85] decomposition method, within NEMD simulations (see Sec. 5 of SM [78]). As can be seen in FIG. 2, the vertical heat transport across both graphene [see FIG. 2(a)] and $h$-BN [see FIG. 2(b)] bilayer interfaces is primarily facilitated by phonons with frequencies below 30 THz, dominated by out-of-plane atomic motion [see FIG. 2(c) and FIG. 2(d)]. The application of in-plane strain significantly reduces the phonon density of states resulting in an overall reduction of thermal transport. Similar results are obtained for twist deformations (see Fig. S6 in Sec. 3 of SM [78]). We note that the variation of the integrated frequency resolved thermal conductance with strain and twist deformations agrees well with the bilayer ITC values obtained from the NEMD simulations (see Fig. S7 in Sec. 5 of SM [78]).

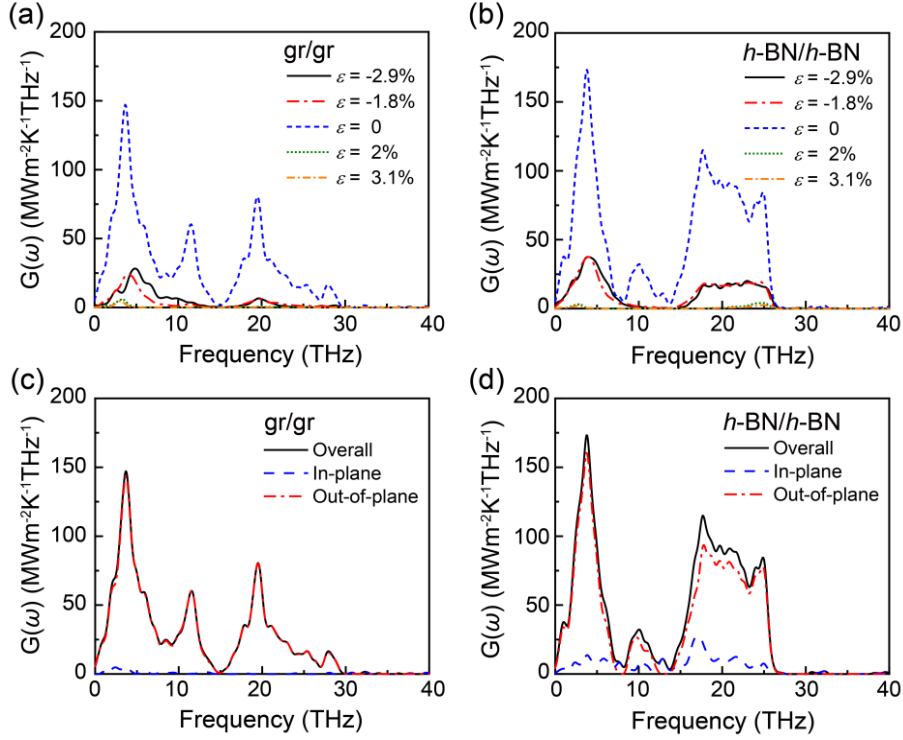

FIG. 2. Frequency resolved thermal conductance ($G(\omega)$) of homogeneous graphene (a) and $h$-BN (b) stacks, calculated for in-plane strain values of $\varepsilon = -2.9\%$ (black), $-1.8\%$ (red), 0 (blue), 2% (green), and 3.1% (orange). Contributions of in-plane (blue dotted line) and out-of-plane (red dash-dotted line) components to overall $G(\omega)$ (black solid line) are presented in panels (c) and (d) for relaxed aligned homogeneous graphene and $h$-BN bilayers, respectively.

A microscopic phenomenological picture should be based on the nature of interlayer interactions within the stack, which are determined by a delicate balance between long-range van der Waals attraction, electrostatic interactions, and short-range Pauli repulsions. While the first two are delocalized in nature, the latter is highly localized and hence influences vertical heat transport the most. The Pauli interaction strength depends mainly on two geometric factors: (i) the vertical distance between neighboring atoms on adjacent layers; and (ii) the local stacking configuration; which, in turn, dictate the interfacial heat transport variation with strain and twist (see Sec. 6 of SM [78]). This intuitive picture allows us to construct a phenomenological model, based on local geometrical characteristics. The model consists of two main ingredients: (i) the local vertical distance between atomic pairs ($d_i$, see Sec. 7 of SM [78]); and (ii) the local registry index (LRI$_i$, see Ref. [86] and Sec. 8 of SM [78]). The latter measures the degree of local interlayer lattice commensurability via projected overlaps between lateral circles, associated with atomic pairs that reside on adjacent layers. The contribution of these two factors to the ITC at any given interface geometry is accounted for via the following formula:

$$\frac{\text{ITC}(\varepsilon,\theta)}{\text{ITC}_0} = \frac{1}{N}\sum_i e^{-\left[a\left(\frac{d_i(\varepsilon,\theta)}{d_0}-1\right)+b\cdot\text{LRI}_i(\varepsilon,\theta)\right]}, \tag{1}$$

where $\text{ITC}_0$ represents the ITC of the aligned ($\theta = 0°$) interface under zero strain ($\varepsilon = 0\%$), $N$ is the total number of atoms in the supercell, and $d_0$ corresponds to the average interlayer distance at the optimal stacking mode ($\varepsilon = 0\%$, $\theta = 0°$). The dimensionless parameters $a$ and $b$ serve as fitting coefficients that quantify the relative contributions of vertical interatomic distance and local atomic stacking to the interlayer heat transport. For the homogeneous bilayer graphene and $h$-BN interfaces the following parameters: $a_{\text{gr}} = a_{h-\text{BN}} = 13.17$, $b_{\text{gr}} = 64.14$, and $b_{h-\text{BN}} = 19.03$, yield good agreements with the heat transport behavior obtained from the more involved atomistic simulations and phonon mode analyses [see FIG. 1(c) and FIG. 1(d) and Fig. S5 in Sec. 3 of SM [78]]. We note that the same value of the $a$ parameter is used for all three interfaces, assuming similar dependence of the vertical heat transport on the interlayer distance. This is supported by the fact that the three systems possess very similar equilibrium interlayer spacings [87]. The differences in the fitted $b$ values indicate the material-specific stacking-dependent contribution to the vertical heat transport.

The same simulation strategy was applied also for the strained and twisted heterogeneous graphene/$h$-BN interface. The heterostructure model consisted of an 8-layer AB stacked graphene slab residing atop an 8-layer AA′ stacked $h$-BN slab. Lateral strain and twist angle effects on the thermal transport were separately considered. Starting from the aligned configuration, the former was applied biaxially to all $h$-BN layers. For the latter, we construct the supercell by identifying superlattice vectors for each of the two layer types that produce negligible strain at the given twist angle [see FIG. 1(a)]. Methodological details regarding the construction of the periodic heterogeneous supercells are provided in Sec. 1 of SM [78].

FIG. 3(a) shows the calculated vertical thermal conductivity through the graphene/$h$-BN stack ($\kappa_\perp$) as a function of strain ($\varepsilon$, black circles). The aligned unstrained graphene/$h$-BN heterostructure is predicted to have a $\kappa_\perp = 0.042 \pm 0.001$ Wm$^{-1}$K$^{-1}$, and the corresponding ITC value of $23 \pm 0.2$ MWm$^{-2}$K$^{-1}$ is well within the range of reported experimental values (~7.4 – 52.2 MWm$^{-2}$K$^{-1}$) [39-41,88]. Up to a strain of 1% (compressive or tensile) the thermal conductivity is only modestly affected. Beyond that, a substantial response of the thermal conductivity to the applied strain is found, where a compressive strain of -3.8% results in a ~98% increase, whereas a 3.2% tensile strain reduces $\kappa_\perp$ by 39%. We note that early indications of the effect of strain in graphene/$h$-BN heterostructures have already been reported in a previous study, where coupled twist and strain deformations resulted in thermal conductivity variations. Here, by separately examining the two contributions, we find that unlike strain deformations, interlayer twisting has only a minor influence on thermal conductivity [see red triangles in FIG. 3(a)].

A similar behavior under strain [see FIG. 3(b)] or twist [see FIG. 3(c)] is exhibited by the (normalized) ITC, as calculated using NEMD simulations (squares), Hessian block diagonalization and Fermi's golden rule (diamonds), or the phenomenological model of Eq. (1) with fitting parameters of $a = 13.17$ and $b = 0$ (stars). As expected, in the case of the incommensurate graphene/$h$-BN heterointerface local stacking effects on the vertical heat transport become negligible, as manifested by the vanishing of the parameter $b$.

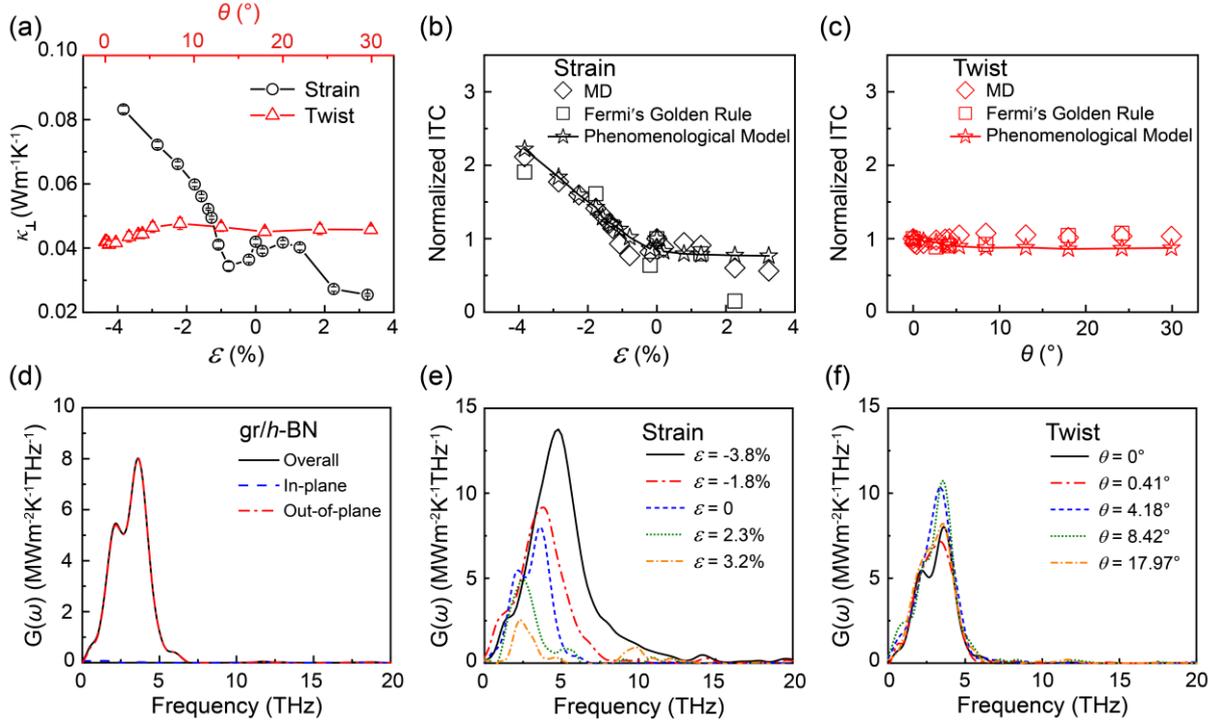

FIG. 3. (a) Strain (black circle, lower horizontal axis) and twist-angle (red triangle, upper horizontal axis) dependence of the vertical thermal conductivity of the graphene/$h$-BN heterojunction. Interfacial heat transfer rates of the graphene/$h$-BN bilayer under various strains (b) and twist angles (c), predicted by NEMD simulation results (black and red diamonds, respectively), are compared to Fermi's golden rule calculations (black and red squares, respectively) and to the phenomenological model predictions (black and red stars, respectively). All values are normalized with respect to that of the corresponding unstrained and aligned interface ($\varepsilon = 0, \theta = 0$). (d) Contributions of in-plane (blue dotted line) and out-of-plane (red dash-dotted line) components to the overall frequency resolved thermal conductance, $G(\omega)$, (black solid line) for the unstrained aligned gr/$h$-BN bilayer. (e), (f) $G(\omega)$ calculated for various strains and twist angles, respectively.

The frequency resolved thermal conductance, $G(\omega)$, of the graphene/$h$-BN bilayer interface, presented in FIG. 3(d), shows that the vertical heat transport is dominated by out-of-plane phonon modes with frequencies below 7 THz. The application of compressive (tensile) strain increases (decreases) the overall density of phonon modes in this range [see FIG. 3(e)] in agreement with the strain-dependence of the heat transport [see FIG. 3(a) and FIG. 3(b)]. Conversely, twist deformations result in minor $G(\omega)$ variations [see FIG. 3(f)], rationalizing the corresponding weak dependence of the heat transport [see FIG. 3(a) and FIG. 3(c)].

The results presented above show that there is a dramatic difference between the vertical thermal conductance response to external mechanical perturbations in homogeneous and heterogeneous layered interfaces of graphene and $h$-BN. For the homogeneous interfaces both lateral strain and interfacial twist induce a significant reduction in thermal conductance. Conversely, while being insensitive to twist deformations, heterogeneous interfaces show either increase or decrease of the vertical thermal conductance under compressive or tensile strains, respectively. Our NEMD atomistic simulation predictions, are rationalized by Fermi's golden rule and density of phonon modes analyses that indicate the central role of vertical phonons and local stacking configurations to the interlayer heat transport behavior. This understanding allows us to construct a simple phenomenological model, based on local interlayer distance and stacking, that captures well the dependence of vertical heat conductance on strain and twist deformations. These insights are valuable for the design and optimization of mechanically tunable van der Waals interfaces for advanced thermal management.

Corresponding Author
*E-mail: huasongqin@xjtu.edu.cn, w.g.ouyang@whu.edu.cn, odedhod@tauex.tau.ac.il

Acknowledgments
W.O. acknowledges support from National Natural Science Foundation of China (Nos. 12472099 and U2441207), and Fundamental Research Funds for the Central Universities (Nos. 2042025kf0050, 2042025kf0013, and 600460100), and the National Key Research and Development Program of China (2025YFF0513900 and 2025YFF0513901). H.Q. acknowledges support from National Natural Science Foundation of China (Grant Nos. 12472108 and T2550247). M.U. acknowledges the financial support of BSF-NSF grant number 2023614. O.H. is grateful for the generous financial support of the Heinemann Chair in Physical Chemistry. Computations were conducted at the Supercomputing Center of Wuhan University, the National Supercomputer TianHe-1(A) Center in Tianjin, and Computing Center in Xi'an.

References

[1]   W. Ouyang, O. Hod, and M. Urbakh, Phys. Rev. Lett. **126**, 216101 (2021).
[2]   D. Mandelli, W. Ouyang, M. Urbakh, and O. Hod, ACS Nano **13**, 7603 (2019).
[3]   B. Hunt *et al.*, Science **340**, 1427 (2013).
[4]   C. R. Dean *et al.*, Nature **497**, 598 (2013).
[5]   B. Lyu *et al.*, Nature **628**, 758 (2024).
[6]   H. Wang, S. Wang, S. Zhang, M. Zhu, W. Ouyang, and Q. Li, Natl. Sci. Rev. **10**, nwad175 (2023).
[7]   G. J. Slotman, M. M. van Wijk, P.-L. Zhao, A. Fasolino, M. I. Katsnelson, and S. Yuan, Phys. Rev. Lett. **115**, 186801 (2015).
[8]   C. R. Woods *et al.*, Nat. Phys. **10**, 451 (2014).
[9]   M. Yankowitz, Q. Ma, P. Jarillo-Herrero, and B. J. LeRoy, Nat. Rev. Phys. **1**, 112 (2019).
[10]  W. Ouyang, D. Mandelli, M. Urbakh, and O. Hod, Nano Lett. **18**, 6009 (2018).
[11]  Y. Song, D. Mandelli, O. Hod, M. Urbakh, M. Ma, and Q. Zheng, Nat. Mater. **17**, 894 (2018).
[12]  D. Mandelli, W. Ouyang, O. Hod, and M. Urbakh, Phys. Rev. Lett. **122**, 076102 (2019).
[13]  K. Huang, H. Qin, S. Zhang, Q. Li, W. Ouyang, and Y. Liu, Adv. Funct. Mater. **32**, 2204209 (2022).
[14]  X. Wu and Q. Han, ACS Appl. Mater. Interfaces **13**, 32564 (2021).
[15]  A. J. Pak and G. S. Hwang, Phys. Rev. Appl. **6**, 034015 (2016).
[16]  X. Liu, G. Zhang, and Y.-W. Zhang, Nano Lett. **16**, 4954 (2016).
[17]  X.-K. Chen, M. Pang, T. Chen, D. Du, and K.-Q. Chen, ACS Appl. Mater. Interfaces **12**, 15517 (2020).
[18]  Z. Yan, L. Chen, M. Yoon, and S. Kumar, Nanoscale **8**, 4037 (2016).
[19]  J. Zhang, Y. Hong, and Y. Yue, J. Appl. Phys. **117**, 134307 (2015).
[20]  W. Ren, S. Lu, C. Yu, J. He, Z. Zhang, J. Chen, and G. Zhang, Appl. Phys. Rev. **10**, 041404 (2023).
[21]  A. H. Castro Neto, F. Guinea, N. M. R. Peres, K. S. Novoselov, and A. K. Geim, Rev. Mod. Phys. **81**, 109 (2009).
[22]  E. Koren, I. Leven, E. Lörtscher, A. Knoll, O. Hod, and U. Duerig, Nat. Nanotechnol. **11**, 752 (2016).
[23]  K. S. Novoselov, A. K. Geim, S. V. Morozov, D. Jiang, M. I. Katsnelson, I. V. Grigorieva, S. V. Dubonos, and A. A. Firsov, Nature **438**, 197 (2005).
[24]  K. S. Novoselov, A. K. Geim, S. V. Morozov, D. Jiang, Y. Zhang, S. V. Dubonos, I. V. Grigorieva, and A. A. Firsov, Science **306**, 666 (2004).
[25]  Y. Zhang, Y.-W. Tan, H. L. Stormer, and P. Kim, Nature **438**, 201 (2005).
[26]  C. R. Dean *et al.*, Nat. Nanotechnol. **5**, 722 (2010).
[27]  A. K. Geim and I. V. Grigorieva, Nature **499**, 419 (2013).
[28]  J. Li *et al.*, Nature **579**, 368 (2020).
[29]  K. S. Novoselov, A. Mishchenko, A. Carvalho, and A. H. Castro Neto, Science **353**, aac9439 (2016).
[30]  L. Wang *et al.*, Science **342**, 614 (2013).
[31]  Z.-G. Chen, Z. Shi, W. Yang, X. Lu, Y. Lai, H. Yan, F. Wang, G. Zhang, and Z. Li, Nat. Commun. **5**, 4461 (2014).
[32]  Y. Yang, J. Ma, J. Yang, and Y. Zhang, ACS Appl. Mater. Interfaces **14**, 45742 (2022).
[33]  B. Lyu *et al.*, Adv. Mater. **34**, 2200956 (2022).
[34]  A. A. Balandin, S. Ghosh, W. Bao, I. Calizo, D. Teweldebrhan, F. Miao, and C. N. Lau, Nano Lett. **8**, 902 (2008).
[35]  S. Ghosh, I. Calizo, D. Teweldebrhan, E. P. Pokatilov, D. L. Nika, A. A. Balandin, W. Bao, F. Miao, and C. N. Lau, Appl. Phys. Lett. **92**, 151911 (2008).
[36]  P. Jiang, X. Qian, R. Yang, and L. Lindsay, Phys. Rev. Mater. **2**, 064005 (2018).
[37]  E. K. Sichel, R. E. Miller, M. S. Abrahams, and C. J. Buiocchi, Phys. Rev. B **13**, 4607 (1976).
[38]  X. Xu *et al.*, Nat. Commun. **5**, 3689 (2014).
[39]  Y. Liu *et al.*, Sci. Rep. **7**, 43886 (2017).

[40] D. B. Brown, T. L. Bougher, X. Zhang, P. M. Ajayan, B. A. Cola, and S. Kumar, Phys. Status Solidi A **216**, 1900446 (2019).
[41] C.-C. Chen, Z. Li, L. Shi, and S. B. Cronin, Appl. Phys. Lett. **104**, 081908 (2014).
[42] W. Ouyang, H. Qin, M. Urbakh, and O. Hod, Nano Lett. **20**, 7513 (2020).
[43] W. Ren, Y. Ouyang, P. Jiang, C. Yu, J. He, and J. Chen, Nano Lett. **21**, 2634 (2021).
[44] M.-L. Lin *et al.*, ACS Nano **12**, 8770 (2018).
[45] Z. Li, J.-M. Lai, and J. Zhang, J. Semicond. **44**, 011902 (2023).
[46] Y. Xiao, J. Liu, and L. Fu, Matter **3**, 1142 (2020).
[47] L. Zhang, Y. Zhong, X. Li, J. H. Park, Q. Song, L. Li, L. Guo, J. Kong, and G. Chen, Nano Lett. **23**, 7790 (2023).
[48] S. E. Kim *et al.*, Nature **597**, 660 (2021).
[49] W. Jiang, T. Liang, J. Xu, and W. Ouyang, Int. J. Heat Mass Transf. **217**, 124662 (2023).
[50] Y. Cheng, Z. Fan, T. Zhang, M. Nomura, S. Volz, G. Zhu, B. Li, and S. Xiong, Mater. Today Phys. **35**, 101093 (2023).
[51] Z. Qin, L. Dai, M. Li, S. Li, H. Wu, K. E. White, G. Gani, P. S. Weiss, and Y. Hu, Adv. Mater. **36**, 2312176 (2024).
[52] F. Eriksson, E. Fransson, C. Linderälv, Z. Fan, and P. Erhart, ACS Nano **17**, 25565 (2023).
[53] W. Jiang, T. Liang, H. Bu, J. Xu, and W. Ouyang, ACS Nano **19**, 16287 (2025).
[54] N. Wei, Y. Zhang, Y. Chen, and Y. Zhang, Int. J. Smart Nano Mater. **11**, 310 (2020).
[55] F. Yang *et al.*, Phys. Rev. Lett. **134**, 146302 (2025).
[56] T. Liang, W. Jiang, K. Xu, H. Bu, Z. Fan, W. Ouyang, and J. Xu, J. Appl. Phys. **138**, 075101 (2025).
[57] B. Xu, M. An, S. Masubuchi, Y. Li, R. Guo, T. Machida, and J. Shiomi, Adv. Funct. Mater. **35**, 2422761 (2025).
[58] H. Bu, W. Jiang, P. Ying, T. Liang, Z. Fan, and W. Ouyang, J. Mech. Phys. Solids **210**, 106540 (2026).
[59] W. Yuan, K. Ueji, T. Yagi, T. Endo, H. E. Lim, Y. Miyata, Y. Yomogida, and K. Yanagi, ACS Nano **15**, 15902 (2021).
[60] W. Ouyang, O. Hod, and R. Guerra, J. Chem. Theory Comput. **17**, 7215 (2021).
[61] W. Ouyang, R. Sofer, X. Gao, J. Hermann, A. Tkatchenko, L. Kronik, M. Urbakh, and O. Hod, J. Chem. Theory Comput. **17**, 7237 (2021).
[62] Z. Feng, Y. Yao, J. Liu, B. Wu, Z. Liu, and W. Ouyang, J. Phys. Chem. C **127**, 8704 (2023).
[63] Y. Yao, B. Wu, Z. Liu, and W. Ouyang, J. Phys. Chem. C **128**, 6836 (2024).
[64] Z. Feng, Z. Lei, Y. Yao, J. Liu, B. Wu, and W. Ouyang, Langmuir **39**, 18198 (2023).
[65] W. Jiang, R. Sofer, X. Gao, A. Tkatchenko, L. Kronik, W. Ouyang, M. Urbakh, and O. Hod, J. Phys. Chem. A **127**, 9820 (2023).
[66] W. Jiang, R. Sofer, X. Gao, L. Kronik, O. Hod, M. Urbakh, and W. Ouyang, J. Phys. Chem. C **129**, 1417 (2025).
[67] Q. Liang, W. Jiang, Y. Liu, and W. Ouyang, J. Phys. Chem. C **127**, 18641 (2023).
[68] W. Ouyang, I. Azuri, D. Mandelli, A. Tkatchenko, L. Kronik, M. Urbakh, and O. Hod, J. Chem. Theory Comput. **16**, 666 (2020).
[69] X. Gao, W. Ouyang, L. Kronik, M. Urbakh, and O. Hod, J. Chem. Phys. **163**, 041001 (2025).
[70] K. Xu *et al.*, Mater. Genome Eng. Adv. **3**, e70028 (2025).
[71] D. W. Brenner, O. A. Shenderova, J. A. Harrison, S. J. Stuart, B. Ni, and S. B. Sinnott, J. Phys.: Condens. Matter **14**, 783 (2002).
[72] A. Kınacı, J. B. Haskins, C. Sevik, and T. Çağın, Phys. Rev. B **86**, 115410 (2012).
[73] Y. Chen, J. Wan, Y. Chen, H. Qin, Y. Liu, Q.-X. Pei, and Y.-W. Zhang, Int. J. Therm. Sci. **183**, 107871 (2023).
[74] H. Qin and W. Jiang, J. Phys. D: Appl. Phys. **57**, 105303 (2024).
[75] W. Jiang, T. Liang, J. Xu, and W. Ouyang, ACS Appl. Mater. Interfaces **17**, 34833 (2025).

[76] W. Yan, J. Liu, W. Ouyang, and Z. Liu, Interdiscip. Mater. **3**, 343 (2024).
[77] A. P. Thompson *et al.*, Comput. Phys. Commun. **271**, 108171 (2022).
[78] See Supplemental Material for methodology, calculation of vertical thermal conductivity, twist-angle-dependent vertical thermal conductivity of homogeneous graphene and *h*-BN stacks, harmonic phonon-phonon couplings across van der Waals interfaces, spectral heat current decomposition, vertical heat transport in different stacking modes, definition of vertical interatomic distances, and local registry index of van der Waals heterointerfaces. The Supplemental Material includes Refs. [79-83].
[79] T. Liang *et al.*, npj Comput. Mater. **12**, 11 (2026).
[80] G. L. Pollack, Rev. Mod. Phys. **41**, 48 (1969).
[81] G. E. Stedman, Am. J. Phys. **39**, 205 (1971).
[82] A. N. Kolmogorov and V. H. Crespi, Phys. Rev. B **71**, 235415 (2005).
[83] K. Sääskilahti, J. Oksanen, S. Volz, and J. Tulkki, Phys. Rev. B **91**, 115426 (2015).
[84] S. Xiong, K. Sääskilahti, Y. A. Kosevich, H. Han, D. Donadio, and S. Volz, Phys. Rev. Lett. **117**, 025503 (2016).
[85] K. Sääskilahti, J. Oksanen, J. Tulkki, and S. Volz, Phys. Rev. B **90**, 134312 (2014).
[86] O. Hod, ChemPhysChem **14**, 2376 (2013).
[87] O. Hod, J. Chem. Theory Comput. **8**, 1360 (2012).
[88] Z. Shang, Z. Guo, Q. Liu, Y. Zhang, X. Wang, M. Zhu, Z. Zhu, S. Qin, and F. Luo, ACS Appl. Mater. Interfaces (2026).